\def\tipo{1}

\def\av#1{\langle#1\rangle}          
\newcommand{\omitit}[1]{}

\ifnum \tipo = 2
\documentclass[twocolumn,showpacs,preprintnumbers,amsmath,amssymb,superscriptaddress]{revtex4}             
\def\figsize{8cm}

\else 
\documentclass[preprint,showpacs,preprintnumbers,amsmath,amssymb,superscriptaddress]{revtex4}
\def\figsize{12cm}

\fi

\usepackage{graphicx}
\usepackage{dcolumn}
\usepackage{bm}
\usepackage{subfigure}

\begin{document}

\preprint{APS/123-QED}

\title{Width of Percolation Transition in Complex Networks}

\author{Tomer Kalisky}
\email{kaliskt@mail.biu.ac.il}
\affiliation{Minerva Center and Department of Physics, Bar-Ilan
  University, 52900 Ramat-Gan, Israel}

\author{Reuven Cohen}
\affiliation{Department of Computer Science and Applied Mathematics,
  Weizmann Institute of Science, Rehovot, Israel}



\date{\today} 

\begin{abstract}
%
  It is known that the critical probability for the percolation
  transition is not a sharp threshold, actually it is a region of
  non-zero width $\Delta p_c$ for systems of finite size.
  Here we present evidence that for complex networks $\Delta p_c \sim
  \frac{p_c}{l}$, where $l \sim N^{\nu_{opt}}$ is the average
  length of the percolation cluster, and $N$ is the number of nodes in
  the network. For Erd\H{o}s-R\'enyi (ER) graphs $\nu_{opt} = 1/3$,
  while for scale-free (SF) networks with a degree distribution $P(k)
  \sim k^{-\lambda}$ and $3<\lambda<4$, $\nu_{opt} =
  (\lambda-3)/(\lambda-1)$.
  We show analytically and numerically that the \textit{survivability}
  $S(p,l)$, which is the probability of a cluster to survive $l$
  chemical shells at probability $p$, behaves near criticality as
  $S(p,l) = S(p_c,l) \cdot exp[(p-p_c)l/p_c]$.  Thus for probabilities
  inside the region $|p-p_c| < p_c/l$ the behavior of the system is
  indistinguishable from that of the critical point.
  
\end{abstract}



\maketitle


\def\figureRecurse{
  \begin{figure}
    \resizebox{\figsize}{!}{\includegraphics{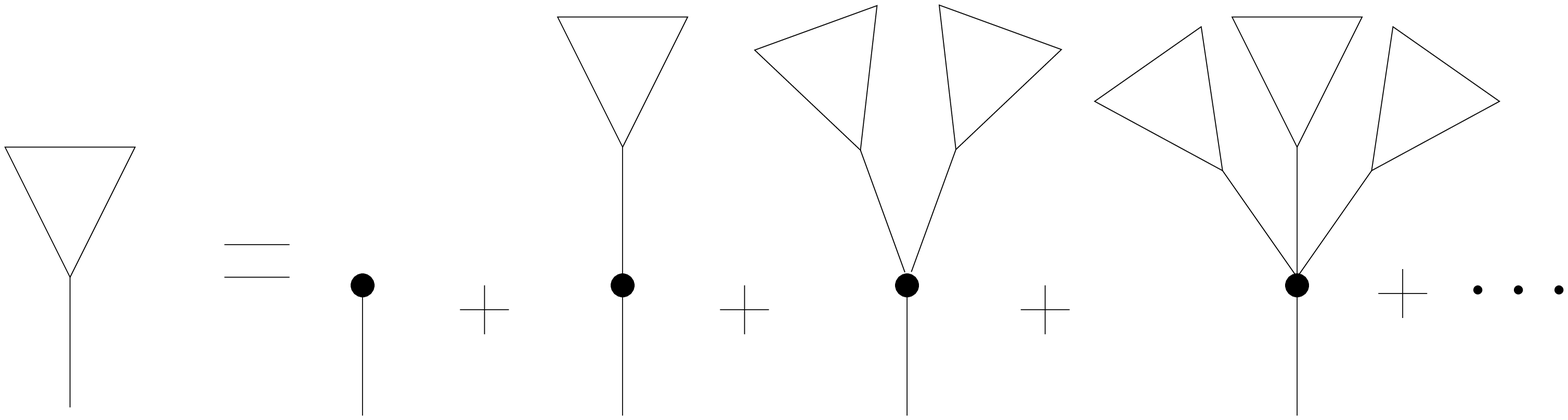}}
    \caption{\label{fig:recurse} A graphical sketch of the recursion
      relation~(\ref{equ:recursive}). Given a certain conduction
      probability $p$, the probability to reach $i$ nodes at layer
      $l+1$ is found by summing up the probabilities to follow a link
      to a node (whose outgoing degree is described by $G_1(x)$) and
      reaching $i$ nodes at layer $l$ from that node.}
  \end{figure}
}

\def\figureERsurvivability{
  \begin{figure}
    \resizebox{\figsize}{!}{\includegraphics{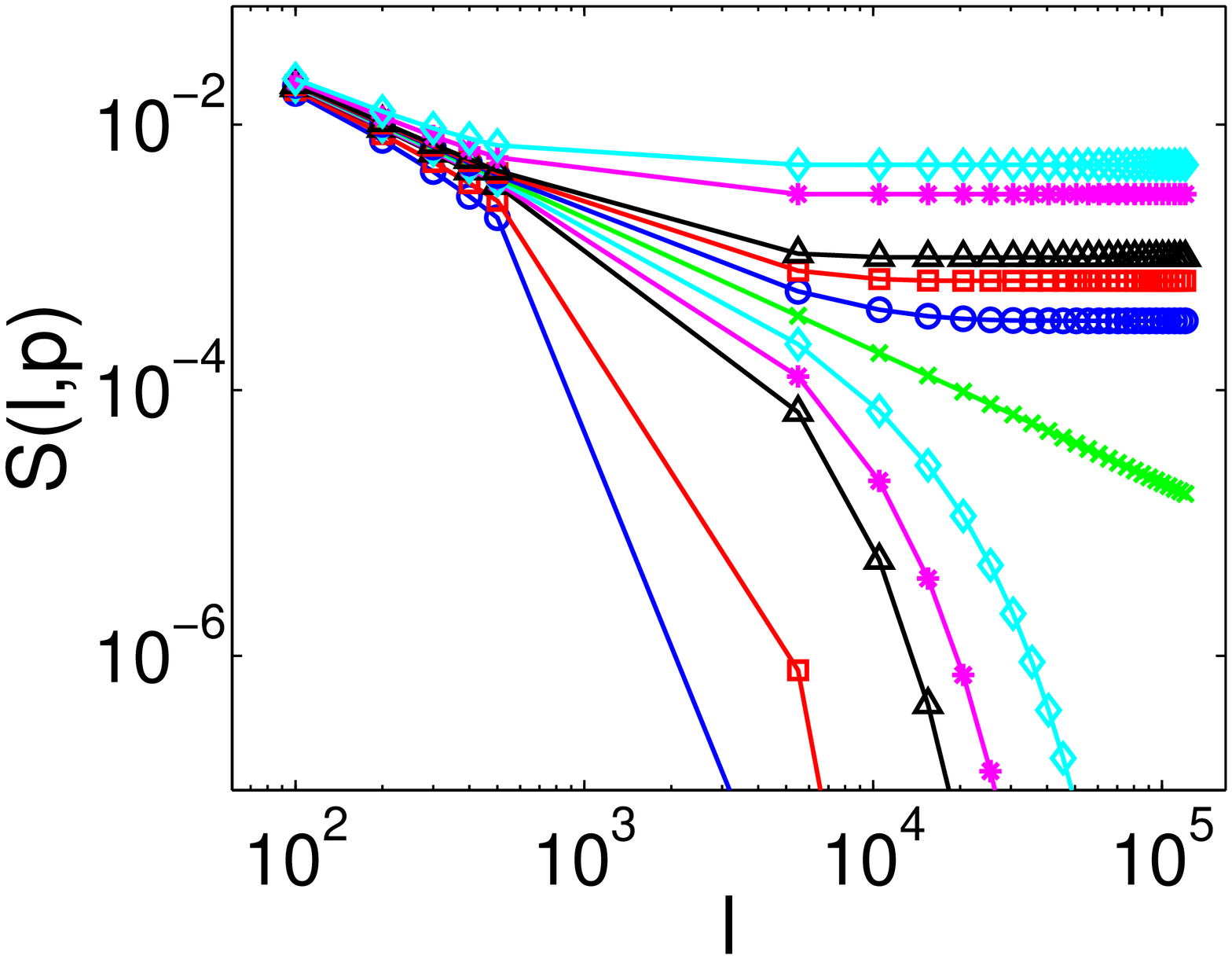}}
    \caption{\label{fig:ERsurvivability} (Color online) The survivability
      $S(p,l)$ for an ER graphs with $\av{k}=5$, numerically
      calculated for different values of $p$: $p_c$, $p_c \pm 5 \cdot
      10^{-4}$, $p_c \pm 3 \cdot 10^{-4}$, $p_c \pm 1 \cdot 10^{-4}$,
      $p_c \pm 6.66 \cdot 10^{-5}$, and $p_c \pm 3.33 \cdot 10^{-5}$.
      For $p=p_c$ the survivability decays to zero according to a
      power law: $S(p_c,l) \sim l^{-1}$. For $p < p_c$, $S(p,l)
      \rightarrow 0$, while for $p > p_c$, $S(p,l) \rightarrow Const$.
      The decay is exponential (to zero or to a constant) according to
      equations~(\ref{equ:ER_survivability})
      and~(\ref{equ:ER_survivability_modified}).}
  \end{figure}
}

\def\figureERsurvivabilityScaled{
  \begin{figure}
    \resizebox{\figsize}{!}{\includegraphics{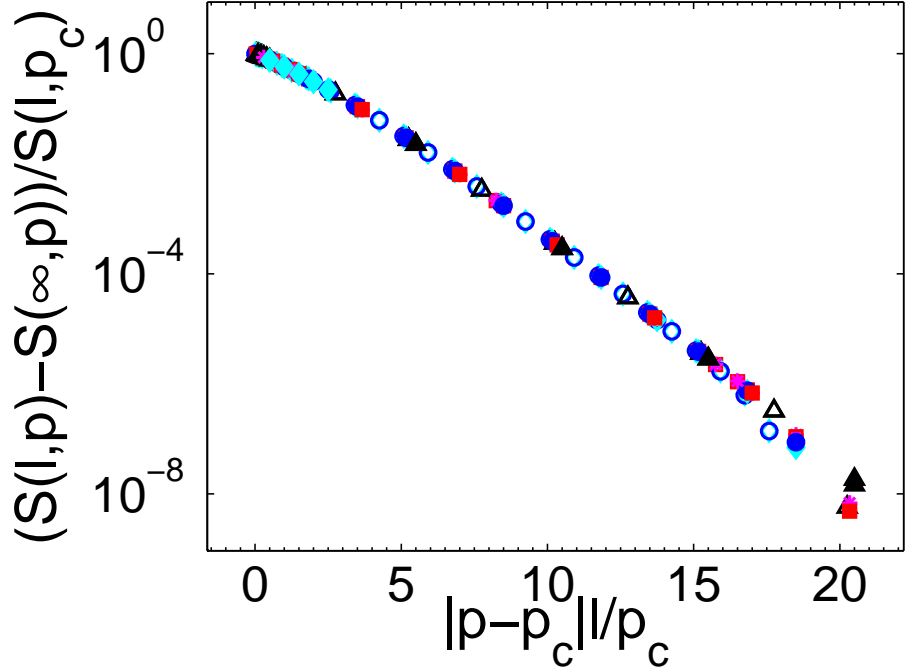}}
    \caption{\label{fig:ERsurvivability_scaled} (Color online) Scaling
      of the survivability for different values of $p$, $l$, and
      $\av{k}$. Shown is $\frac{S(p,l)-S(p,\infty)}{S(p_c,l)}$ vs.
      $\frac{1}{p_c}|p-p_c|l$ for ER graphs with $\av{k}=5$ (unfilled
      symbols) and $\av{k}=10$ (filled symbols). The collapse of all
      curves on an exponential function (for large $l$) shows that
      indeed the scaling relations~(\ref{equ:ER_survivability})
      and~(\ref{equ:ER_survivability_modified}) are correct.}
  \end{figure}
}

\def\figureSFsurvivability{
  \begin{figure}
    \resizebox{\figsize}{!}{\includegraphics{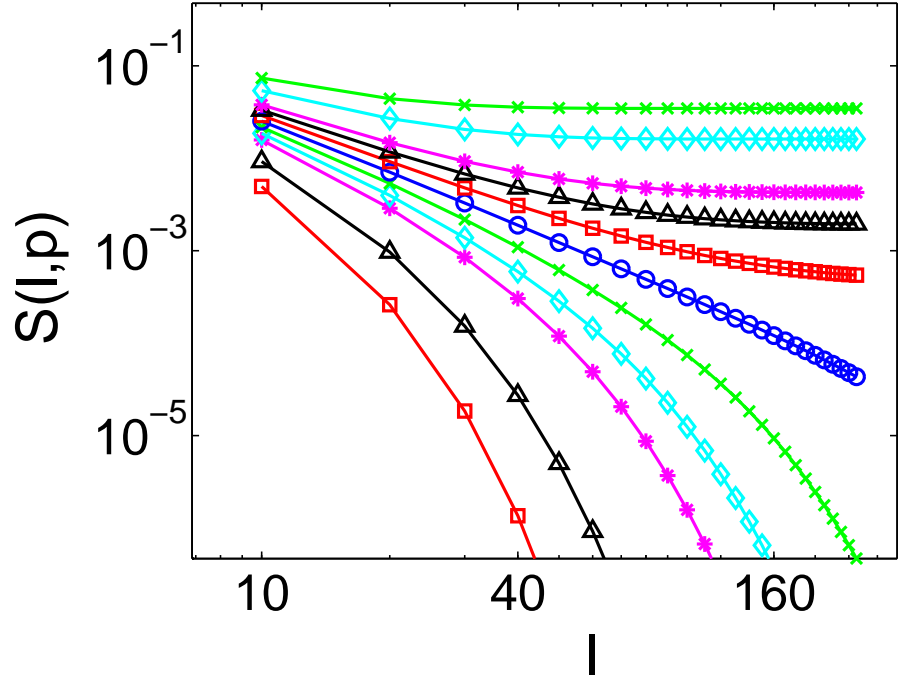}}
    \caption{\label{fig:SFsurvivability} (Color online) The survivability
      $S(p,l)$ for a SF network with $\lambda=3.5$, numerically
      calculated for different values of $p$: $p_c$, $p_c \pm 6 \cdot
      10^{-2}$, $p_c \pm 4 \cdot 10^{-2}$, $p_c \pm 2 \cdot 10^{-2}$,
      $1.33 \cdot 10^{-2}$, and $p_c \pm 6.66 \cdot 10^{-3}$. For
      $p=p_c$ the survivability decays to zero according to a power
      law: $S(p_c,l) \sim l^{-2}$. For $p < p_c$, $S(p,l) \rightarrow
      0$, while for $p > p_c$, $S(p,l) \rightarrow Const$.  The decay
      is exponential (to zero or to a constant) according to
      equations~(\ref{equ:ER_survivability})
      and~(\ref{equ:ER_survivability_modified}).}
  \end{figure}
}

\def\figureSFsurvivabilityScaled{
  \begin{figure}
    \resizebox{\figsize}{!}{\includegraphics{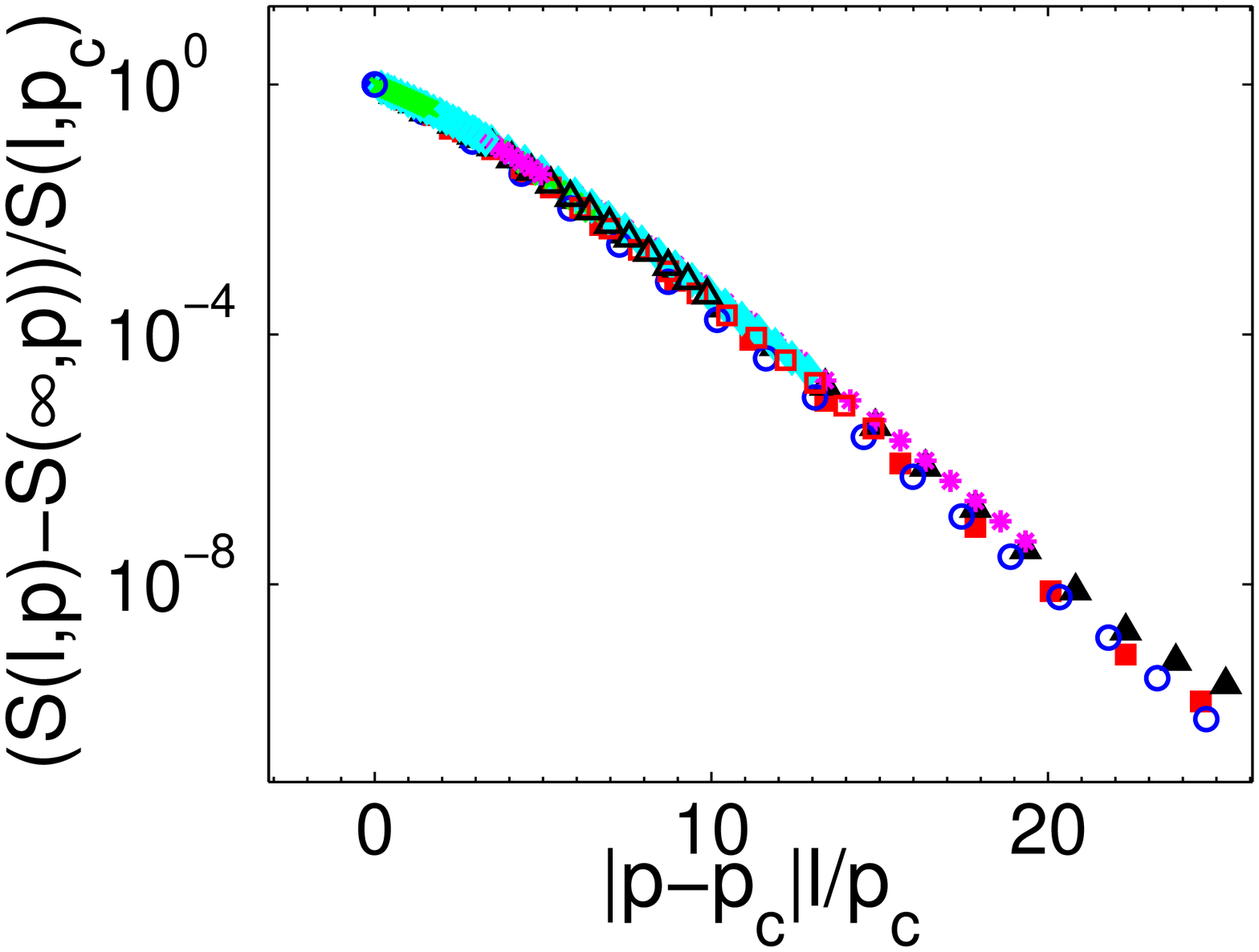}}
    \caption{\label{fig:SFsurvivability_scaled} (Color online) Scaling
      of the survivability for different values of $p$, $l$, and
      $\lambda$. Shown is $\frac{S(p,l)-S(p,\infty)}{S(p_c,l)}$ vs.
      $\frac{1}{p_c}|p-p_c|l$ for SF graphs with $\lambda=3.5$ (filled
      symbols) and $\lambda=5$ (unfilled symbols). Due to numerical
      difficulties only curves with $p<p_c$ are
      shown~\protect\footnote{The exact enumeration method is limited
        here to small chemical distances $l$ due to the upper cutoff
        of the scale-free distribution.}. All curves collapse on an
      exponential function according to
      relation~(\ref{equ:ER_survivability}).}
  \end{figure}
}

\section{\label{sec:Introduction}Introduction:}

Recently the subject of networks has received much attention. It was
realized that many systems in the real world, such as the Internet,
can be successfully modeled as networks. Other examples include social
networks such as the web of social contacts, and biological networks
such as the protein interaction network and metabolic
networks~\cite{Barabasi-2002:Linked,Dorogovtsev-Mendes-2003:From_Biological_Nets_to_the_Internet,vespignani-pastor-satorras-2004:evolution_and_structure}.
The problem of percolation on networks has also been studied
extensively (e.g.~\cite{Barabasi-Albert-2002:statistical-mechanics}).
Using percolation theory we can describe the resilience of the network
to breakdown of sites or
links~\cite{newman-callaway-2000:networks_robustness,Cohen-Erez-Ben_Avraham-Havlin-2000:Resilience},
epidemic spreading~\cite{Cohen-ben-avraham-Havlin-2003:immunization},
and the properties of the optimal path in a highly network with highly
fluctuating weights on the
links~\cite{Braunstein-Buldyrev-Cohen-Havlin-Stanley-2003:Optimal}.

A typical percolation system consists of a $d$-dimensional grid of
length $L$, in which the nodes or links are removed with some
probability $1-p$, or are considered ``conducting'' with probability
$p$
(e.g.~\cite{Bunde-Havlin-1996:Fractals-and-Disordered-Systems,stauffer-aharoni-1992:percolation_theory}).
Below some critical probability $p_c$ the system becomes disconnected
into small clusters, i.e., it becomes impossible to cross from one
side of the grid to the other by following the conducting links.
Percolation is considered a geometrical phase transition exhibiting
universality, critical exponents, upper critical dimension at $d=6$
etc. It was noted by Conigilio~\cite{coniglio-1982:cluster_structure}
that for systems of finite size $L$ the transition from connected to
disconnected state has a width $\Delta p_c \sim \frac{1}{L^{1/\nu}}$,
where $\nu$ is a critical exponent related to the correlation length.

Percolation on networks was studied also from a mathematical point of
view~\cite{Erdos-Renyi-1960:Evolution,bollobas-2001:random_graphs,Barabasi-Albert-2002:statistical-mechanics}.
It was found that in Erd\H{o}s-R\'enyi (ER) graphs with an average
degree $\av{k}$ the percolation threshold is: $p_c =
\frac{1}{\av{k}}$. Below $p_c$ the graph is composed of small clusters
(most of them trees). As $p$ approaches $p_c$ trees of increasing
order appear. At $p=p_c$ a giant component emerges and loops of all
orders abruptly appear. Nevertheless, for graphs of finite size $N$ it
was found that the percolation threshold has a finite width $\Delta
p_c \sim \frac{1}{N^{1/3}}$~\cite{bollobas-2001:random_graphs},
meaning that all attributes of criticality are present in the system
in the range $[p_c - \Delta p_c,p_c + \Delta p_c]$. For example: The
number of loops is negligible below $p_c + \Delta p_c$~\footnote{O.
  Riordan and P.~L.~Krapivsky, private communication.}.

In this paper we study the \textit{Survivability} of the network near
the critical threshold. The survivability $S(p,l)$ is defined to be
the probability of a connected cluster to ``survive'' up to $l$
chemical shells in a system with conductance probability
$p$~\cite{tzschichholz-bunde-havlin-1989:loopless} (i.e the
probability that there exists at least one node at chemical distance
$l$ from a randomly chosen node on the same cluster). At the critical
point $p_c$, the survivability decays as a power-law: $S(p_c,l) \sim
l^{-x}$, where $x$ is a universal exponent. Below $p_c$ the
survivability decays exponentially to zero, while above $p_c$ it decays
(exponentially) to a constant. Here we will derive analytically and
numerically the functional form of the survivability above and below
the critical point. We will show that near the critical point $S(p,l)
= S(p_c,l) \cdot exp[(p-p_c)l/p_c]$. Thus, given a system with a
maximal chemical length $l$, for probabilities inside the range
$|p-p_c| < \frac{p_c}{l}$ the behavior of the system is
indistinguishable from that of the critical point. Hence we get
$\Delta p_c \sim \frac{p_c}{l}$.

The maximal chemical length $l$ at criticality is actually the length
of the percolation cluster, which was found to be: $l \sim
N^{\nu_{opt}}$ where $N$ is the number of nodes in the network. For
Erd\H{o}s-R\'enyi (ER) graphs $\nu_{opt} = 1/3$, while for scale-free
(SF) networks with a degree distribution $P(k) \sim k^{-\lambda}$ and
$3<\lambda<4$, $\nu_{opt} =
(\lambda-3)/(\lambda-1)$~\cite{Braunstein-Buldyrev-Cohen-Havlin-Stanley-2003:Optimal}.

\section{\label{sec:ER_graphs}Erd\H{o}s-R\'enyi Graphs:}

Consider an ER graph with a mean degree $\av{k}$, and each link
having a probability $p$ to conduct. We define $N_l(x) = n_0 + n_1 x +
n_2 x^2 + n_3 x^3 + ...$ to be the generating function of the number
of sites that exists on layer (i.e. chemical shell) $l$ starting from
a random node on the graph (for a conduction probability $p$).

The generating function for the degree distribution of a randomly
chosen node in the network is $G_0(x) = \sum P(k) \cdot x^k$ and the
generating function for the number of links emerging from a node
\textit{reached by following a randomly chosen link} is $G_1(x) =
\Sigma \frac{1}{\av{k}} kP(k) \cdot
x^{k-1}$~\cite{Newman-Strogatz-Watts-2001:Random}. Taking into
account the probability $p$ for conduction, we have:
\begin{equation}
  \label{equ:G_1}
  G_1(x) = 1-p + p \sum \frac{1}{\av{k}} kP(k) \cdot x^{k-1}
  .
\end{equation}

We can now write the following recursive
relation~\cite{Kalisky-Cohen-ben-Avraham-Havlin-2004:Tomography}:
\begin{equation}
  \label{equ:recursive}
  N_{l+1}(x) = G_1(N_l(x))
  ,
\end{equation}
which means that the probability $n^{(l+1)}_i$ for having $i$ nodes at
layer $l+1$ is composed of the probability of reaching a vertex by
following a link, and reaching $i$ nodes at layer $l$ by following all
branches emerging from that vertex - see sketch in
Fig.~\ref{fig:recurse}.

\ifnum\tipo=2
\figureRecurse
\fi

It can be seen that $N_l(0) = n_0$ is the probability that there are
$0$ nodes at layer $l$, i.e., the probability to die before layer $l$.
Thus $\epsilon_l = 1 - N_l(0)$ is the probability to survive up to layer
$l$.
From~(\ref{equ:recursive}) we have: 
\begin{equation}
  N_{l+1}(0) = G_1(N_l(0))
\end{equation}
\begin{equation}
  1-\epsilon_{l+1} = G_1(1-\epsilon_l)
\end{equation}
\begin{equation}
  \label{equ:recursive_1}
  1-\epsilon_{l+1} = 1 - p + p \sum \frac{1}{\av{k}} kP(k) (1-\epsilon_l)^{k-1}
\end{equation}
For ER graphs $G_0(x) = G_1(x) = e^{{\av{k}(x-1)}}$ (for $p=1$), Thus:
\begin{equation}
  1-\epsilon_{l+1} = 1 - p + p e^{\av{k} (1 - \epsilon_l  - 1)} = 1 - p + p e^{-\av{k} \epsilon_l}
\end{equation}
\begin{equation}
  \label{equ:recursive_ER}
  \epsilon_{l+1} = p - p e^{-\av{k} \epsilon_l}
\end{equation}
Setting $\delta \equiv p - p_c$, where $p_c = \frac{1}{\av{k}}$, and
expanding by series we get:
\begin{equation}
  \epsilon_{l+1} = p_c + \delta - (p_c + \delta) \left(1 - \av{k} \epsilon_l + \frac{1}{2} \av{k}^2 \epsilon_l^2 - ... \right)
\end{equation}
Leaving only expressions up to second order in $\delta$ and
$\epsilon_l$ (we assume that $p < p_c$ and thus $\epsilon_l \ll 1$ for
large $l$) we get:
\begin{equation}
  \epsilon_{l+1} \approx \epsilon_l - \frac{1}{2} \av{k} \epsilon_l^2 + \delta \av{k} \epsilon_l
\end{equation}
\begin{equation}
  \label{equ:recursive_ER_near_pc}
  \frac{d \epsilon_{l}}{dl} \approx - \frac{1}{2} \av{k} \epsilon_l^2 + \frac{\delta}{p_c} \cdot \epsilon_l
\end{equation}
At criticality, $\delta = 0$ and the solution to this equation is:
$\epsilon_l \sim
l^{-1}$~\cite{Kalisky-Cohen-ben-Avraham-Havlin-2004:Tomography}. The
additional term suggests the following solution near criticality:
$\epsilon_l \sim l^{-1} \cdot \exp{\left( \frac{1}{p_c} \delta l
  \right)}$
\footnote{Indeed, solving equation~(\ref{equ:recursive_ER_near_pc})
  taking $\delta \ll 1$ and the initial condition $\epsilon_{l=0}=1$
  we get: $\epsilon_l = \frac{2}{\av{k}} l^{-1} \exp{\left(
      \frac{1}{p_c} \delta l \right)}$.}.

In terms of survivability this can be written as:
\begin{equation}
  \label{equ:ER_survivability}
  S(p,l) = S(p_c,l) \cdot \exp \left(\frac{1}{p_c}(p-p_c)l \right)
  .
\end{equation}
In order to check this result we numerically solved the survivability
$S(p,l)$ near $p_c$ according to the exact enumeration method
presented
in~\cite{Braunstein-Buldyrev-Sreenivasan-Cohen-Havlin-Stanley-2004:the_optimal}~\footnote{This
  method assumes that near the critical point there is a negligible
  number of loops and thus the network behaves similar to a cayley
  tree with the same degree distribution $P(k)$ as the ER network.}.
Fig.~\ref{fig:ERsurvivability} shows the survivability $S(p,l)$ for
different values of $p$. For $p = p_c$ the survivability decays as a
power law, while above and below there is an exponential decay, either
to zero (for $p<p_c$) or to a constant (for $p>p_c$).
Fig.~\ref{fig:ERsurvivability_scaled} shows that all curves of the
survivability $S(p,l)$ from Fig.~\ref{fig:ERsurvivability} can be
rescaled such that they all collapse. Moreover, scaled survivabilities
from all different graphs with different values of $\av{k}$ (i.e.,
different values of $p_c$) also collapse on the same curve.
However, equation~(\ref{equ:ER_survivability}) is true only below the
percolation threshold where there is no giant component. Above the
percolation threshold there is an exponential decay to a non-zero
constant, and the generalized expression is:
\begin{equation}
  \label{equ:ER_survivability_modified}
  S(p,l) = S(p_c,l) \cdot \exp \left(- \frac{1}{p_c} |p-p_c| l \right) + pP_{\infty}
  ,
\end{equation}
Where $pP_{\infty}$ is the probability for a randomly chosen site to be
inside the percolation cluster~\footnote{$S(p,l \rightarrow \infty)$
  is the probability that is we start from a randomly chosen
  conducting site, we will survive an infinite chemical distance.
  This equals the probability $p$ that the randomly chosen site is
  conducting, multiplied by the probability $P_{\infty}$ that it
  resides in the giant component.}.
Indeed, setting $\epsilon_{l+1} = \epsilon_l$ in
equation~(\ref{equ:recursive_ER}) the resulting ``steady state''
solution is
$pP_{\infty}$~\cite{bollobas-2001:random_graphs}\footnote{$P_{\infty}$
  obeys the transcendental equation: $P_{\infty} = 1-e^{-\av{k} p
    P_{\infty}}$.}.

\ifnum\tipo=2
\figureERsurvivability
\figureERsurvivabilityScaled
\fi

\section{\label{sec:SF_graphs}Scale-Free Graphs:}

Scale-free graphs can be taken to have a degree distribution of the
form $P(k) = c k^{-\lambda}$ where $c \approx (\lambda-1)
m^{\lambda-1}$~\cite{Cohen-Erez-Ben_Avraham-Havlin-2000:Resilience}.
In order to solve equation~(\ref{equ:recursive_1}) we have to
evaluate:
\begin{equation}
  G_1(1-\epsilon_l) = \frac{1}{\av{k}} \sum kP(k) (1-\epsilon_l)^{k-1}
\end{equation}
Expanding by powers of $\epsilon_l$, and inserting $P(k)=c
k^{-\lambda}$ with $3<\lambda<4$, we
get~\cite{cohen-havlin-2005:complex_networks}:
\begin{equation}
  \sum kP(k) (1-\epsilon_l)^{k-1} \approx \av{k} - \av{k(k-1)} \epsilon_l + \frac{c}{2} \Gamma(4-\lambda) \epsilon_l^{\lambda-2}
\end{equation}
Thus equation~(\ref{equ:recursive_1}) becomes:
\begin{equation}
  1-\epsilon_{l+1} \approx 1 - p +
  \frac{p}{\av{k}} \left( \av{k} - \av{k(k-1)} \epsilon_l + \frac{c}{2} \Gamma(4-\lambda) \epsilon_l^{\lambda-2} \right)
\end{equation}
Taking $p = p_c + \delta$:
\begin{equation}
  1-\epsilon_{l+1} \approx 1 - (p_c+\delta) +
  \frac{p_c+\delta}{\av{k}} \left( \av{k} - \av{k(k-1)} \epsilon_l + \frac{c}{2} \Gamma(4-\lambda) \epsilon_l^{\lambda-2} \right)
  .
\end{equation}
Substituting
$p_c=\frac{\av{k}}{\av{k(k-1)}}$~\cite{Cohen-Erez-Ben_Avraham-Havlin-2000:Resilience}
we get:
\begin{equation}
  1-\epsilon_{l+1} \approx 1 - \epsilon_l + p_c \frac{c}{2 \av{k}} \Gamma(4-\lambda) \cdot \epsilon_l^{\lambda-2}
  - \frac{1}{p_c} \cdot \delta \epsilon_l + \frac{c}{2 \av{k}} \Gamma(4-\lambda) \cdot \delta \epsilon_l^{\lambda-2}
\end{equation}
Setting $A \equiv p_c \frac{c}{2 \av{k}} \Gamma(4-\lambda)$ we get:
\begin{equation}
  \epsilon_{l+1} - \epsilon_l \approx -A \cdot \epsilon_l^{\lambda-2} + \frac{1}{p_c} \cdot \delta \epsilon_l
  -  \frac{A}{p_c} \cdot \delta \epsilon_l^{\lambda-2}
\end{equation}
\begin{equation}
  \epsilon_{l+1} - \epsilon_l \approx -A \cdot \epsilon_l^{\lambda-2}
  + \frac{1}{p_c} \cdot \delta \left( \epsilon_l  - A \cdot \epsilon_l^{\lambda-2} \right)
  .
\end{equation}
For large $l$, $\epsilon_l \ll 1$, and taking into account that
$\lambda-2 > 1$ we have $\epsilon_l^{\lambda-2} \ll \epsilon_l$.
Therefore:
\begin{equation}
  \label{equ:recursive_SF_near_pc}
  \frac{d \epsilon_l}{dl} \approx -A \cdot \epsilon_l^{\lambda-2} + \frac{1}{p_c} \cdot \delta \epsilon_l 
  .
\end{equation}
For $\delta = 0$ the solution is $\epsilon_l \sim l^{-x}$ with $x =
\frac{1}{\lambda-3}$~\cite{Kalisky-Cohen-ben-Avraham-Havlin-2004:Tomography}.
The additional term suggests the following solution near criticality:
$\epsilon_l \sim l^{-x} \cdot \exp{\left( \frac{\delta l}{p_c}
  \right)}$
\footnote{Solving equation~(\ref{equ:recursive_SF_near_pc}) with
  $\delta \ll 1$ and the initial condition $\epsilon_{l=0}=1$ we get:
  $\epsilon_l = \frac{1}{( (\lambda-3)A )^{ 1/(\lambda-3) }}
  l^{-1/(\lambda-3)} \exp{\left( \frac{1}{p_c} \delta l \right)}$.}.
A similar form can be found for $\lambda >
4$~\footnote{In this range is the behavior is similar to ER
  graphs~\cite{Cohen-Ben_Avraham-Havlin-2002:critical_exponents}.}.
The scaling form for SF networks is also confirmed by numerical
simulations as shown in Figures~\ref{fig:SFsurvivability}
and~\ref{fig:SFsurvivability_scaled}.

\ifnum\tipo=2
\figureSFsurvivability
\figureSFsurvivabilityScaled
\fi

\section{\label{sec:summary}Summary and Conclusions}

The scaling form of the survivability near the critical probability
obeys the following scaling relation (for $p<p_c$):
\begin{equation}
  \label{equ:survivability_fluct}
  S(p,l) = S(p_c,l) \cdot \exp \left(\frac{p-p_c}{\Delta p_c} \right)
  .
\end{equation}
Where $\Delta p_c = \frac{p_c}{l}$. Given a system with a maximal
chemical length $l$, for all values of conductivity $p$ inside the
range $[p_c - \Delta p_c,p_c + \Delta p_c]$ the survivability behaves
similar to the power law $S(p_c,l) \sim l^{-x}$ found at criticality.
Thus, the width of the critical threshold is $\Delta p_c =
\frac{p_c}{l}$.

To summarize, we have shown analytically and numerically the the
survivability in ER and SF graphs scales according to
equations~(\ref{equ:ER_survivability})
and~(\ref{equ:ER_survivability_modified}) near the critical point.
This implies that the width of the critical region in networks of
finite size scales as $\Delta p_c = \frac{p_c}{l}$, where $l$ is the
chemical length of the percolation cluster. For ER graphs, $l \sim
N^{1/3}$, while for SF networks with $3<\lambda<4$, $l \sim
N^{(\lambda-3)/(\lambda-1)}$.

\section*{Acknowledgments}
We thank the ONR, the Israel Science Foundation and the Israeli Center
for Complexity Science for financial support. We thank E.~Perlsman, S.
Sreenivasan, Lidia A.~Braunstein, Sergey V. Buldyrev, Shlomo Havlin,
H. Eugene Stanley, Y. Strelniker, Alexander Samukhin, O. Riordan and
P.~L.~Krapivsky for useful discussions.
\omitit{Lidia A.~Braunstein thanks the ONR - Global for financial
  support.}

\appendix

\bibliography{surv} 
  
  
  

\ifnum\tipo=1
\figureRecurse
\figureERsurvivability
\figureERsurvivabilityScaled
\figureSFsurvivability
\figureSFsurvivabilityScaled
\fi

\end{document}